\documentclass[12pt]{article}
\usepackage{amssymb}
\usepackage{graphicx,color}
\usepackage{amsmath}
\usepackage{enumerate}
\usepackage{array}

\usepackage{epsfig}
\usepackage[english]{babel}
\usepackage{latexsym}

\usepackage{bm}
\def\boldnabla{{\bm \nabla}}
\def\eps{\varepsilon}

\def\mv{{\bm v}}

\def\J{\mathcal{J}}

\def\R{{\scriptscriptstyle R}}

\def\dRM{{\mathrm d}}

\def\mk{{\bm k}}

\def\mx{{\bm x}}

\def\bmu{{\bm u}}

\def\boldnabla{{\bm \nabla}}
\allowdisplaybreaks

\newcommand{\fp}[2]{FP$^{\textrm{#1}}_{#2}$}

\def\sod{{\leavevmode\setbox1=\hbox{d}%
\hbox to 1.05\wd1{d\kern-0.4ex{\char039}\hss}}}
\def\sot{{\leavevmode\setbox1=\hbox{t}%
\hbox to \wd1{t\kern-0.5ex{\char039}\hss}}}
\def\sol{l\kern-0.3ex\raise0.1ex\hbox{'}\kern-0.10ex}
\def\soL{L\kern-0.8ex\raise0.1ex\hbox{'}\kern0.1ex}
\begin{document}
\title{Gribov process advected by the synthetic compressible velocity
ensemble: Renormalization Group Approach}

\author{N.~V.~Antonov$^{1}$, M. Hnatich$^{2,3,4}$, A.~S.~Kapustin$^{1}$,\\ T. Lu\v{c}ivjansk\'y$^{3,5}$, L.~Mi\v{z}i\v{s}in$^{3}$}

\maketitle\mbox{ }
\\
 $^{1}$ Department of Theoretical Physics, St. Petersburg 
University,\\Ulyanovskaya 1, St. Petersburg, Petrodvorets, 198504 Russia,\\
 $^{2}$ Institute of Experimental
Physics, Slovak Academy of Sciences, Watsonova 47, 040 01
Ko\v{s}ice, Slovakia,\\
$^{3}$ Faculty of Sciences, P.J. \v{S}af\'arik
University, Moyzesova 16, 040 01 Ko\v{s}ice, Slovakia,\\
$^4$ Bogoliubov Laboratory of Theoretical Physics, JINR, 141980 Dubna, Moscow Region, Russia,\\
$^{5}$ Fakult\"at f\"ur Physik, Universit\"at Duisburg-Essen, D-47048 Duisburg, Germany\\

\begin{abstract}
The direct bond percolation process (Gribov process) is studied in the presence of
 irrotational velocity fluctuations with long-range correlations. The perturbative renormalization group 
 is employed in order to analyze the effects of finite correlation time on the long-time behavior 
 of the phase transition between an active and an absorbing
state. The calculation is performed to the one-loop order. 
Stable fixed points of the renormalization group and their regions of stability are
 obtained within the three-parameter $(\eps,y,\eta)$-
expansion. Different regimes corresponding to the rapid-change limit and frozen velocity field
are discussed.
\end{abstract}
{\section{Introduction} \label{sec:intro}}
For a long time non-equilibrium continuous phase transitions \cite{Hin06} have been an object
of intense research activity. Underlying dynamic laws are responsible
for diverse behavior with respect to their equilibrium counterparts.
One of the most prominent example is the directed bond 
 percolation \cite{Stauffer,HHL08} process, also known as Schl\"ogl first reaction \cite{Schlogl,Grassberger82}. 
 In particle physics this process has been introduced by Gribov \cite{Gribov} in order
 to explain hadron interactions at very high
 energies (Reggeon field theory) \cite{Cardy}. Further it can serve for a description of
 stochastic reaction-diffusion processes on a lattice \cite{Hinrichsen} and spreading 
 of infection diseases \cite{Janssen81} among others.  
 

It has been known that phase transitions are quite sensitive with respect
to additional disturbances such as quenched disorder \cite{Janssen96} or long-range
interactions \cite{Hinrichsen}. From practical point of view
this might be a reason why there are not so many experimental
realizations for the percolation process  \cite{RRR03,TKCS07}.
Majority of realistic reaction-diffusion processes occur in some fluid environment, e.g., vast majority
 of chemical reactions is realized at finite temperature, which is inevitable accompanied with the 
 presence of thermal fluctuations.
In this paper we assume that the effect of environment can be simulated by advective velocity 
fluctuations \cite{FGV01}. 
Dynamics of the fluids is governed by the Navier-Stokes equation \cite{Landau_fluid}.
A general solution of these equations still remains an open question
 \cite{Frisch,Monin}. Kraichnan model appears
 as a more tractable problem. In this model velocity field is
 assumed to obey a Gaussian distribution law with prescribed  statistical properties \cite{FGV01,Ant99}.
 Though at a first sight too
 oversimplified with respect to the realistic flows, it nevertheless captures an essential
 physical information about advection processes \cite{FGV01}. Moreover some
 properties as intermittency are even more pronounced than for the Navier Stokes equation itself.
 
Recently, there has been increased interest in different advection problems in 
compressible turbulent flows \cite{Benzi09,Pig12,Volk14,depietro15}. These studies show that
compressibility plays an important role for population dynamics or chaotic mixing of colloids.
 In this work we  
 consider a generalization of the original Kraichnan model proposed in \cite{Ant00}.  There
advection-diffusion problem of non-interacting admixture  was studied
in the presence of velocity field with finite correlation in time and compressibility taken into account.
 Our main motivation is to use this model and determine what influence it has
 on the critical properties of the directed bond percolation process. 
   We note that in our model there is no backward influence of percolating
 field on the velocity fluctuations, i.e., our model corresponds to
 the passive advection of the scalar quantity.
As initial steps in this direction have already been undertaken \cite{AntKap08,AntKap10,Ant11,DP13}, our main
aim here is to elucidate in detail the
 differences between incompressible and compressible velocity field. The main
 theoretical tool is the field-theoretic approach \cite{Vasiliev} with subsequent Feynman diagrammatic
 technique and renormalization group (RG) approach, which allows us to determine large-scale
 behavior.
 
The paper is organized as follows. In 
 Sec.~\ref{sec:model}, we introduce a
field-theoretic version of the problem. In Sec.~\ref{sec:renorm} we 
 describe main steps of the
 perturbative RG procedure. In Sec.~\ref{sec:stable} we present an 
 analysis of possible regimes involved in the model. 
 We analyze numerically
 and to some extent analytically  fixed points' structure.
 In Sec.~\ref{sec:conclu} we give a concluding summary. 
{\section{Field-theoretic model} \label{sec:model}}
The effective field-theoretic action \cite{JanTau04} for directed percolation can be obtained
either from the Langevin formulation or via reaction-diffusion 
scheme employing Doi formalism \cite{Doi}. However, at the very end one arrives at
 the same De Dominicis-Janssen action \cite{Janssen76,deDom76,Janssen79}
\begin{equation}
  \J_{ \text{per}}[\tilde{\psi},\psi] =  
  \tilde{\psi}[
  \partial_t + D_0(\tau_0 -\nabla^2)
  ]\psi  + 
   \frac{D_0\lambda_0}{2} [\psi-\tilde{\psi}]\tilde{\psi}\psi,
  \label{eq:act_per}
\end{equation}
where all irrelevant contributions from the RG point of view have been neglected.
Field $\psi$ corresponds to the fluctuating density of percolating agents,
$\tilde{\psi}$ stands for auxiliary Martin-Siggia-Rose field (MSR), 
$\partial_t = \partial / \partial t$ is
the time derivative, $\nabla^2$ is  the Laplace operator, $D_0$ 
is the diffusion constant, $g_0$ is the coupling constant and $\tau_0$ measures
 a deviation from the threshold value for injected probability. It can be thought
 as an analog to the temperature  variable in the standard $\varphi^4-$theory 
 \cite{JanTau04,Zinn}.
 For the future RG use  we have extracted  a 
 dimensional part from the interaction terms  in the action (\ref{eq:act_per}).
In this paper we use a condensed notation in which expressions are viewed
as matrices or vectors with respect to component indices, the spatial variable 
$x$ and the time variable $t$, respectively. 
For example, the first term in the action (\ref{eq:act_per}) actually reads
 \begin{equation}
   \int \dRM t \int \dRM^{d} x\mbox{ } \tilde{\psi}(t,x)
   \partial_t \psi(t, x),
 \end{equation}
 where $d$ is the dimensionality of the $\mx$ space.
Here and henceforth 
we distinguish between
unrenormalized (with a subscript ``0'') quantities and renormalized terms
(without a subscript ``0'').

The next step consists of an incorporation of the velocity fluctuations into
the model. The 
 standard route \cite{Landau_fluid} is based on the replacement 
\begin{equation}   
   \partial_t \rightarrow \partial_t +(v_i\partial_i),
   \label{eq:lagr_der}
\end{equation}   
where the summation over the spatial index $i$ is implied.
 In accordance with \cite{Ant99,Ant00} we assume that
 the velocity field is a random Gaussian variable with zero mean and 
 a translationally invariant correlator \cite{Ant00} given
 in the Fourier representation 
\begin{equation}
  \langle v_i v_j \rangle_0 (\omega,k)
  = [P_{ij}^{k} + \alpha Q_{ij}^{k}]
  \frac{g_{10} u_{10} D_0^3 k^{4-d-y-\eta}}{\omega^2 + u_{10}^2 D_0^2 (k^{2-\eta})^2}.
  \label{eq:kernelD}
\end{equation}
Here, $P_{ij}^k = \delta_{ij}-k_ik_j/k^2$ is a transverse  and $Q_{ij}^k=k_ik_j/k^2$ a longitudinal
projection
operator, respectively. Further $k=|\mk|$ and a positive parameter $\alpha>0$ can be interpreted as the simplest possible
deviation \cite{AdzAnt98} from the incompressibility condition
$\partial_i v_i = 0$.
The incompressible case, $\alpha=0$, has been analyzed in previous 
works \cite{AntKap08,Ant11,DP13}.
The coupling constant $g_{10}$ and the exponent $y$ describe the equal-time velocity
correlator or, equivalently, the energy spectrum \cite{Frisch,Ant99,Ant00} of the velocity
fluctuations. The constant $u_{10}>0$ and the exponent $\eta$ are related
to the characteristic frequency $\omega \simeq u_{10} D_0 k^{2-\eta}$ of the mode with
wavelength $k$.

The kernel function for the correlator (\ref{eq:kernelD}) has been chosen in a universal
 form
 and as such it contains different limits: rapid-change model, frozen
 velocity ensemble and turbulence advection (see \cite{Ant99,Ant00}).
 In this paper our main goal is to analyze the case
 of purely potential (irrotational) velocity field.
To this end one more rescaling of the variable $g_1$ according to
\begin{equation}
   \alpha g_1 \rightarrow g_1, \quad \alpha\rightarrow\infty.
   \label{eq:rescale}
\end{equation}
is needed. Then, in the perturbation theory we effectively work with the following velocity propagator
\begin{equation}
  \langle v_i v_j \rangle_0 (\omega,k) \rightarrow
  k_i k_j
  \frac{g_{10} u_{10} D_0^3 k^{2-d-y-\eta}}{\omega^2 + u_{10}^2 D_0^2 (k^{2-\eta})^2},
  \label{eq:potential_v}
\end{equation}
where we have relabeled $g_1\alpha \rightarrow g_1$.
In the functional language the Gaussian nature of velocity field reveals in
the following quadratic action
\begin{align}
  \J_{\text{vel}}[v] = \frac{1}{2}
  v D^{-1} v,
  \label{eq:act_vel}
\end{align}
 which has to be added to the complete field-theoretical functional.
\section{Renormalization group analysis \label{sec:renorm}}
In order to apply the dimensional regularization for an evaluation 
of renormalization constants, an analysis of possible superficial
divergences must be performed.
For translationally invariant
systems, it  is sufficient  \cite{Zinn,Amit} to analyze 1-particle irreducible (1PI)
graphs only.
In contrast to static models, dynamical models  \cite{Vasiliev,Tauber2014} 
contain two independent scales: a frequency scale 
 $d^\omega_Q $ and a momentum scale $d^k_Q$ for each quantity $Q$.
The corresponding dimensions are found using the 
standard normalization conditions 
\begin{align}
  d_k^k = - d^k_x =1,\quad
  d^k_\omega & = d_t^k = 0, \quad 
  d_k^\omega = d^\omega_x = 0,\quad
  d^\omega_\omega = -d_t^\omega = 1
  \label{eq:def_normal}
\end{align}
together with a condition field-theoretic action to be a dimensionless quantity.
Using values $d^\omega_Q$ and $d_Q^k$,
the total canonical dimension $d_Q$,
\begin{equation}
   d_Q = d_Q^k + 2d_Q^\omega
\end{equation}
can be introduced,
whose precise form is obtained from a comparison of IR most
relevant terms ($\partial_t$ must scale as $\boldnabla^2$) in the action (\ref{eq:act_per}).
The total dimension $d_Q$ for the dynamical models
plays the same role as the conventional (momentum) dimension does in static problems.
Dimensions of all quantities for the model are summarized in Table \ref{tab:canon}.
To retain the standard notation we have introduced $\eps$ via relation $d=4-\eps$.
It follows that the model is logarithmic (when coupling constants
are dimensionless) at $\eps = y = \eta = 0$, and the UV divergences are
in principle realized as poles in these parameters. For the RG analysis it is
of crucial importance that the couplings become logarithmic at the same time. 
Otherwise, one would have to discard IR irrelevant ones and some scaling regimes
will be absent.
\begin{table}[h!]
 \centering
 \setlength\extrarowheight{5pt}
\begin{tabular}{| c | c | c | c | c| c | c | c | c | }
  \hline
  $Q$ & $\psi,\tilde{\psi}$ & ${\mv}$ & $D_0$ & $\tau_0$ & $g_{10}$ & $\lambda_0 $  
  & $u_{10}$  & $u_{20},a_0,\alpha$
 \\  \hline
  $d_Q^k$ & $d/2$ & $-1$ & $-2$ & $2$ & $y$ & $\eps/2$
  & $\eta$  & $0$ 
  \\  \hline
  $d^\omega_Q$ & 0 & $1$ & $1$ & $0$ & $0$ & $0$
  & $0$  & $0$ 
  \\  \hline
  $d_Q$ & $d/2$ & $1$ & $0$ & $2$ & $y$ & $\eps/2$
  & $\eta$  & $0$ 
  \\ \hline    
\end{tabular}
 \caption{Canonical dimensions of the bare fields and bare parameters 
	  for the total field-theoretic action given by the sum of actions for
	  percolation process (\ref{eq:act_per}), velocity field (\ref{eq:act_vel}) and 
	  advection process (\ref{eq:inter_act}).  }
  \label{tab:canon}
\end{table}
The total canonical dimension of an arbitrary $1-${\it irreducible} Green function
is given by the relation 
\begin{equation}
  d_\Gamma = d^k_\Gamma + 2 d^\omega_\Gamma = d + 2 -
  \sum_\varphi N_\varphi d_\varphi, \varphi\in\{\tilde{\psi}, \psi, \mv \}.
  \label{eq:def_dim}
\end{equation}
The total dimension $d_\Gamma$ in the logarithmic theory is a formal degree of the 
UV divergence $\delta_\Gamma = d_\Gamma |_{\eps=y=\eta=0}$. 
Superficial UV divergences, whose removal requires counterterms, could be
present only in those functions $\Gamma$ for which $\delta_\Gamma$ is
a non-negative integer \cite{Vasiliev}.

\begin{table}[h!]
  \centering
  \setlength\extrarowheight{2pt}
  \begin{tabular}{|c|c|c|c|c|c|c|c|c|}
    \hline
    $\Gamma_{1-ir}$ &  $\Gamma_{\tilde{\psi} \psi} $& $\Gamma_{\tilde{\psi} \psi \mv }$
                  & $\Gamma_{\tilde{\psi}^2 \psi} $ & $ \Gamma_{\tilde{\psi}\psi^2} $
                  & $\Gamma_{\tilde{\psi}\psi \mv^2}$                                    
   \\
   \hline
    $d_\Gamma$ & $2$ & $1$ & $\eps/2$ & $\eps/2$ & $0$
               \\
    \hline
     $\delta_\Gamma$ & $2$ & $1$ & $0$ & $0$ & $0$
               \\ \hline
  \end{tabular}
  \caption{Canonical dimensions for the (1PI) divergent Green functions of the model.}
  \label{tab:canon_green}
\end{table}
Dimensional analysis should be augmented by certain additional considerations.
In dynamical models with MSR response fields \cite{Tauber2014}, all
the 1-irreducible diagrams without the
fields $\tilde{\psi}$ vanish, and it is sufficient to consider functions with 
$N_{\tilde{\psi}} \ge 1$. As was shown in \cite{AntKap10} the 
rapidity symmetry $\psi(t)\rightarrow -\tilde{\psi}(-t), \tilde{\psi}\rightarrow-\psi(-t)$
requires also inequality $N_{\psi} \ge 1$ to hold. Using these considerations
together with relation (\ref{eq:def_dim}), possible UV divergent structures
are expected only for the 1PI Green functions listed in
Table \ref{tab:canon_green}.
	
In what follows we employ the perturbative RG approach, which allows us to 
calculate universal quantities in formal series in small parameter of theory.
In contrast to the standard $\varphi^4$-theory our model contains three
small expansion parameters $(\eps,\eta,y)$. Also we would like to make the 
following remark. The real expansion parameters in a perturbative sense
are the charges $g_1$ and $g_2=\lambda^2$ only (the latter fact is a consequence
of rapidity symmetry). The parameters $u_1$ and $\alpha$ correspond to
the non-perturbative quantities, and there is no physical restriction
on their values. Therefore one can study also a limiting case such as
$u_{10} \rightarrow 0$ or $u_{10} \rightarrow \infty$.
	
Before we embark on results of the RG approach, let us first discuss in detail
profound differences caused by compressibility and lack of Galilei invariance \cite{AAK94,ABG96,AV97} in our model.
As shown in \cite{AntKap10} instead of relation (\ref{eq:lagr_der}) the following
   replacement is necessary
\begin{equation}
  \partial_t \rightarrow \partial_t +(v_i\partial_i)+a_0 (\partial_i v_i),
  \label{eq:subs}
\end{equation}
where $a_0$ is an additional positive parameter, whose significance can be explained as follows.
For pure advection-diffusion problem \cite{Ant00} the choice $a_0=1$ corresponds to the conserved
quantity $\psi$ (density),  whereas for $a_0=0$ auxiliary field $\tilde{\psi}$ is
conserved. For the whole model nor $\psi$ neither $\tilde{\psi}$ is conserved, hence both
 fields $\tilde{\psi}$ and $\psi$ are fluctuating quantities and RG procedure
will give a birth to both counterterms
$\tilde{\psi} (v_i \partial_i) \psi $ and $\tilde{\psi} \partial_i(v_i\psi)$.
In the language of Feynman diagrams let us consider a one-loop expansion of 1PI function ${\tilde{\psi} {\psi} \mv}$ that
can be formally written as
\begin{align}
  \Gamma_{\tilde{\psi} {\psi} \mv}
  & =  
  -ip_j Z_4 - ia q_j Z_5 +
  \raisebox{-4.25ex}{ \epsfysize=1.75truecm \epsffile{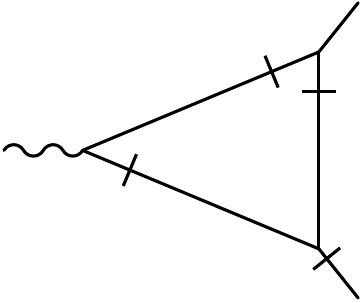}} 
  + 
  \raisebox{-4.25ex}{ \epsfysize=1.75truecm \epsffile{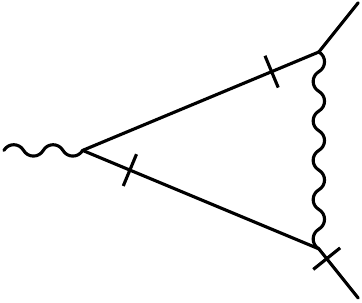}}, 
  \label{eq:exp_ppv}
\end{align}
where $p$ is a momentum of the field $\psi$ and $q$ of the field $\mv$, respectively.
 A direct calculation shows that a divergent part of the first graph is
\begin{equation}
   \frac{i\lambda_0^2}{4d(2\pi)^d\eps}[2p_j-q_j].
   \label{eq:div_triple}
\end{equation}
Let us consider such graph as a subgraph in some high-order loop for an incompressible case.
In this case all velocity propagators are proportional to the transverse projector and after contraction
with (\ref{eq:div_triple}) the compressible part evidently drops out. However, in our model the 
 velocity propagator (\ref{eq:kernelD}) contains also a longitudinal part.
Moreover in contrast to Kraichnan model, also the second graph in (\ref{eq:exp_ppv}) is  divergent
 (due to the finite correlation in time the graph does not contain a closed loop of retarted propagators \cite{Vasiliev,Tauber2014}).
Symmetries play a fundamental role in physics. In turbulent problems  Galilei invariance \cite{Vasiliev} is of prominent
importance. It describes an invariance with respect to transformations
$\varphi\rightarrow\varphi_v$ given by
\begin{align} 
  \varphi_v(x)& = \varphi(x_v) - v(t), &\varphi'_v(x)& = \varphi'(x_v),\quad x \equiv (t,\mx),\nonumber\\
  x_v &\equiv (t,\mx+\bmu(t)),
  &\bmu(t)& =  \int_{-\infty}^t \dRM t' \mv(t'). 
\end{align}
From Table \ref{tab:canon_green} it follows that possible superficial divergences may appear
also in the structure $\Gamma_{\tilde{\psi}\psi \mv^2}$. Aforementioned Galilei invariance
restricts presence of such term. In our model however such term must be taken into
account and considered as a new interaction parameter. The one-loop expansion for this 
function using only cubic interactions reads
\begin{align}
  \Gamma_{\tilde{\psi}{\psi} \mv^2}
  & =  \frac{u_{2}}{D}\delta_{ij} Z_8 +
  \raisebox{-4.25ex}{ \epsfysize=1.75truecm \epsffile{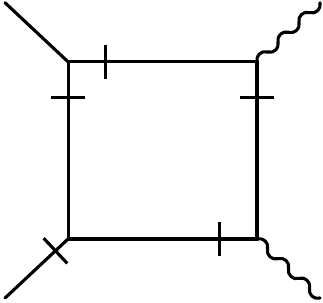}} 
   +   
  \raisebox{-4.25ex}{ \epsfysize=1.75truecm \epsffile{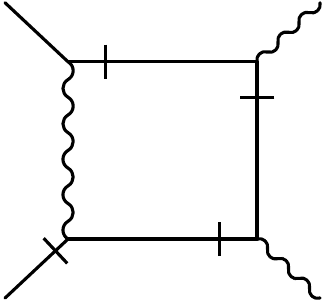}} 
  + \frac{1}{2}
  \raisebox{-4.25ex}{ \epsfysize=1.75truecm \epsffile{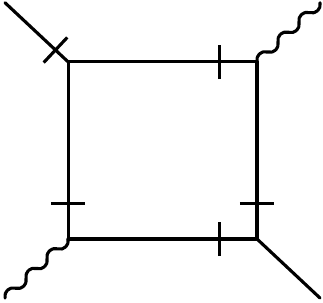}}.  
  \label{eq:exp_ppvv}
\end{align}	
 A straightforward calculation shows that the first and the third graph cancel each other, whereas
second graph gives a non-zero contribution. This is again a consequence of finite correlation time, which
 precludes appearance of closed loops of retarted propagators.
To conclude the field-theoretic action given by the sum of (\ref{eq:act_per}) and (\ref{eq:act_vel}) has to be augmented by the following 
part describing the advection interaction
\begin{align}
  \J_{\text{adv}}[\varphi] = 
  -\frac{u_{20}}{2D_0} \tilde{\psi} \psi
  v^2  +  \tilde{\psi} (v_i\partial_i) \psi 
  +a_0 \tilde{\psi} (\partial_i v_i)\psi.
  \label{eq:inter_act}
\end{align}
The field theoretic action ${\J}=\J_{\text{per}} + \J_{\text{vel}} + \J_{\text{adv}}$ is amenable to the Feynman diagrammatic technique with
 the subsequent use of perturbative RG approach. Note that the added interaction term $\tilde{\psi}\psi v^2$ directly leads
 to additional $5$ Feynman diagrams for 1PI function (\ref{eq:exp_ppvv})
\begin{align}
  \raisebox{-4.25ex}{ \epsfysize=1.75truecm \epsffile{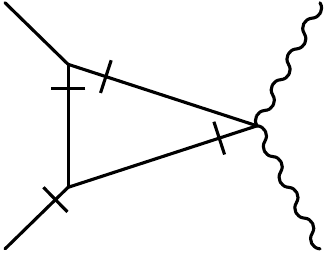}} 
   +   
  \raisebox{-4.25ex}{ \epsfysize=1.75truecm \epsffile{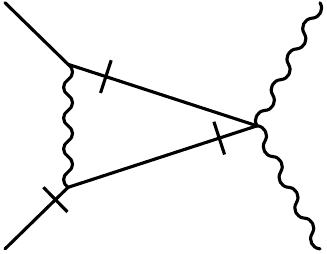}} 
  + 
  \raisebox{-4.25ex}{ \epsfysize=1.75truecm \epsffile{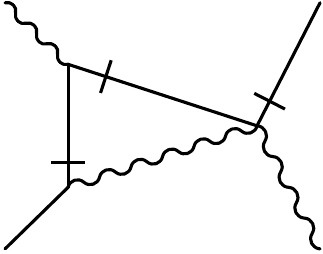}}
  +
  \raisebox{-4.25ex}{ \epsfysize=1.75truecm \epsffile{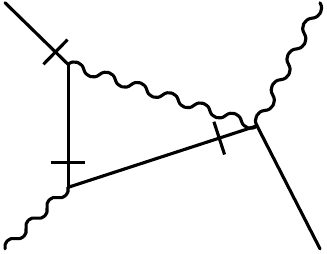}} 
   +   
  \raisebox{-4.25ex}{ \epsfysize=1.75truecm \epsffile{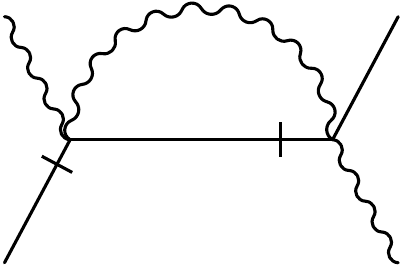}}
  \label{eq:ppvv_new}
\end{align}	 
that have to be taken into account.
The multiplicative renormalization can be achieved through following renormalization
prescription 
\begin{align}
   \label{eq:RGconst}
   &D_0 = D Z_D, &\tau_0& = \tau Z_\tau + \tau_c,
     &a_0& = a Z_a,  &g_{20}&=g_2 \mu^{\eps} Z_{g_2},
     \nonumber \\ 
   &g_{10} = g_{1} \mu^{y} Z_{g_1}, &u_{10}& = u_1 \mu^\eta Z_{u_1},
   &\lambda_0& = \lambda \mu^{\eps/2} Z_\lambda, 
   &u_{20}& = u_2 Z_{u_2}, 
   \nonumber \\
   &\tilde{\psi} = Z_{\tilde{\psi}}\tilde\psi_{\R}, &\psi& = Z_\psi \psi_{ \R},
   &\mv& = Z_v{\mv}_{\R}, 
\end{align}
where $\mu$ is the reference mass scale in the MS scheme~\cite{Zinn}. Note that
the term $\tau_c$ is a non-perturbative effect \cite{JanTau04}, which
is not captured by the  dimensional regularization. Physically it describes fluctuation-induced
shift of critical probability $\tau_0$. 
{\section{IR stable regimes} \label{sec:stable}}
The large scale behavior with respect to spatial and time variables is
governed by the attractive IR stable fixed points $g^*$. Here and henceforth 
the asterisk refers to a coordinate of the fixed point (FP).
Their coordinates are
determined from the zeros of RG flow equations \cite{Vasiliev,Zinn,Amit} 
\begin{align}
  &\beta_{g_1} (g^{*}) =\beta_{g_2} (g^{*})= \beta_{u_1}
  (g^{*})=\beta_{u_2} (g^{*})=\beta_{a} (g^{*})=0.
  \label{eq:gen_beta}
\end{align}
The eigenvalues of the matrix of first derivatives $\Omega=\{\Omega_{ij}\}$ 
determine whether given FP is IR stable. Such points are proper canditates for macroscopic regimes and
thus can be observed experimentally.
 The matrix $\Omega$ is defined as
\begin{equation}
   \Omega_{ij} = \frac{\partial \beta_i}{\partial g_j},\quad
   i,j\in\{g_1,g_2,u_1,u_2,a\}.
   \label{eq:matrix}
\end{equation} 
The explicit form of beta functions follows from (\ref{eq:RGconst}) using definition $\beta_g = \mu \partial_{\mu} g |_{0}$ and 
 a straightforward calculation yields
\begin{align}
   \beta_{g_1} &= g_1 (-y + 2\gamma_D-2\gamma_v), 
      &\beta_{g_2}& =  g_2 (-\eps -\gamma_{g_2}),
      &\beta_{a} &= - a \gamma_{a},
      \nonumber \\
   \beta_{u_1} &= u_1(-\eta +\gamma_D), 
   &\beta_{u_2}&= - u_2 \gamma_{u_2}, 
  \label{eq:beta_functions}
\end{align}
where $\gamma_x \equiv \mu\partial_\mu \ln  Z_x |_0$ are
the anomalous dimensions \cite{Vasiliev}.  In the 1-loop approximation they are given
by the following expressions
\begin{align}
 \gamma_{g_1} & = -\frac{g_1 u_1(1-2u_2)}{2(1+u_1)^2}
 - \frac{g_2}{4},\quad
 \gamma_D  =  \frac{g_1}{4(1+u_1)}\biggl[
      \frac{u_1-1}{u_1+1}+\frac{4 a(1-a)}{(1+u_1)^2}
      \biggl] + \frac{g_2}{8} , \nonumber\\
  \gamma_a & = (1-2a)\biggl[
      \frac{g_1(1-a)}{2(1+u_1)^3} + \frac{g_1 u_2(u_1-1)}{4a(1+u_1)^2} 
       + \frac{g_2}{8a}
  \biggl] , \nonumber\\
  \gamma_{u_2} & = 
    \frac{g_1(1-2u_2)}{4(1+u_1)}\biggl[
    \frac{u_1-1}{u_1+1} + \frac{2 a(1-a)}{u_2(1+u_1)^2}
    \biggl]
  -\frac{g_2}{8}, \nonumber\\
  \gamma_{g_2} & = 
   \frac{g_1}{1+u_1}\biggl[
   \frac{(1-2a)^2}{2} + \frac{1-3a(1-a)}{1+u_1}
   + \frac{2a(1-a)u_1}{(1+u_1)^2} 
   \biggl]
   -\frac{3g_2}{2}, \nonumber\\
   \gamma_{\tau} & = -\frac{g_1}{4(1+u_1)^2}
   \biggl(
   u_1-1 + \frac{4a(1-a)}{1+u_1}\biggl)
   - \frac{3g_2}{8}.
    \label{eq:gen_anom_charges} 
\end{align}
The fields $\psi,\tilde{\psi}$ and $\mv$ also have to be renormalized and therefore
corresponding anomalous dimensions are nontrivial
\begin{align}
  \gamma_{\psi} & = \frac{g_1 }{2(1+u_1)^2}\biggl[
  -a(1-a) + (1+u_1)(2a-1)
  \biggl] - \frac{g_2}{8}, \nonumber\\ 
  \gamma_{\tilde{\psi}} & =
  \frac{g_1}{2(1+u_1)^2}\biggl[
  -a(1-a) + (1+u_1)(1-2a)
  \biggl] - \frac{g_2}{8}, \nonumber\\ 
  \gamma_{\mv} & =  \frac{g_1}{4(1+u_1)^2}\biggl[
     \frac{4a(1-a)}{1+u_1}-1 \biggl] + \frac{g_1 u_1 u_2}{2(1+u_1)^2}.
 \label{eq:gen_anom_fields}
\end{align}
In order to simplify the analysis it is convenient to introduce new charges $g_1',u_1'$ and $a'$
\begin{equation}
  \frac{g_1}{1+u_1} = g'_1 \rightarrow g_1, \quad \frac{1}{1+u_1} = u_1'\rightarrow
  u_1, \quad  (1-2a)^2 = a' \rightarrow a.
  \label{eq:new_charges}
\end{equation}
In new variables the rapid change model corresponds to the choice $u_1=0$, whereas
frozen velocity field to $u_1=1$. The final expressions for $\beta$-functions read
\begin{align}
  \beta_{g_1} & =  \frac{g_1}{8} \biggl\{- 8 y + 8 \eta (1 - u_1) + 
                 2 g_1 (1- u_1) [1 + u_1 (2 + u_1(a- 1)) - 4 u_2] + g_2(1+u_1) \biggl\},\nonumber \\
  \beta_{g_2} & = \frac{g_2}{4}  \biggl\{-4 \eps  + g_1 u_1 (2 u_1 - 3) - a g_1 (2 + u_1 + 2 u_1^2) + 6 g_2\biggl\},\nonumber \\
  \beta_{u_1} & = \frac{u_1 (1-u_1)}{8}   \biggl\{8 \eta  + 2 g_1 [u_1 (2 + u_1(-1 + a))-1] -g_2\biggl\},\nonumber \\
  \beta_{u_2} & = \frac{1}{8}  \biggl\{ g_1 (1 - 2 u_2) [u_1^2 (a-1) - 2 u_2 + 4 u_1 u_2] + g_2 u_2\biggl\},\nonumber \\
  \beta_{a} & = \frac{a}{2}  \biggl\{ g_1 [u_1 (u_1 - a u_1 - 4 u_2) + 2 u_2] + g_2\biggl\}.
  \label{eq:final_beta}
\end{align}

\begin{figure}[h!]
  \centering
  \includegraphics[width=7cm]{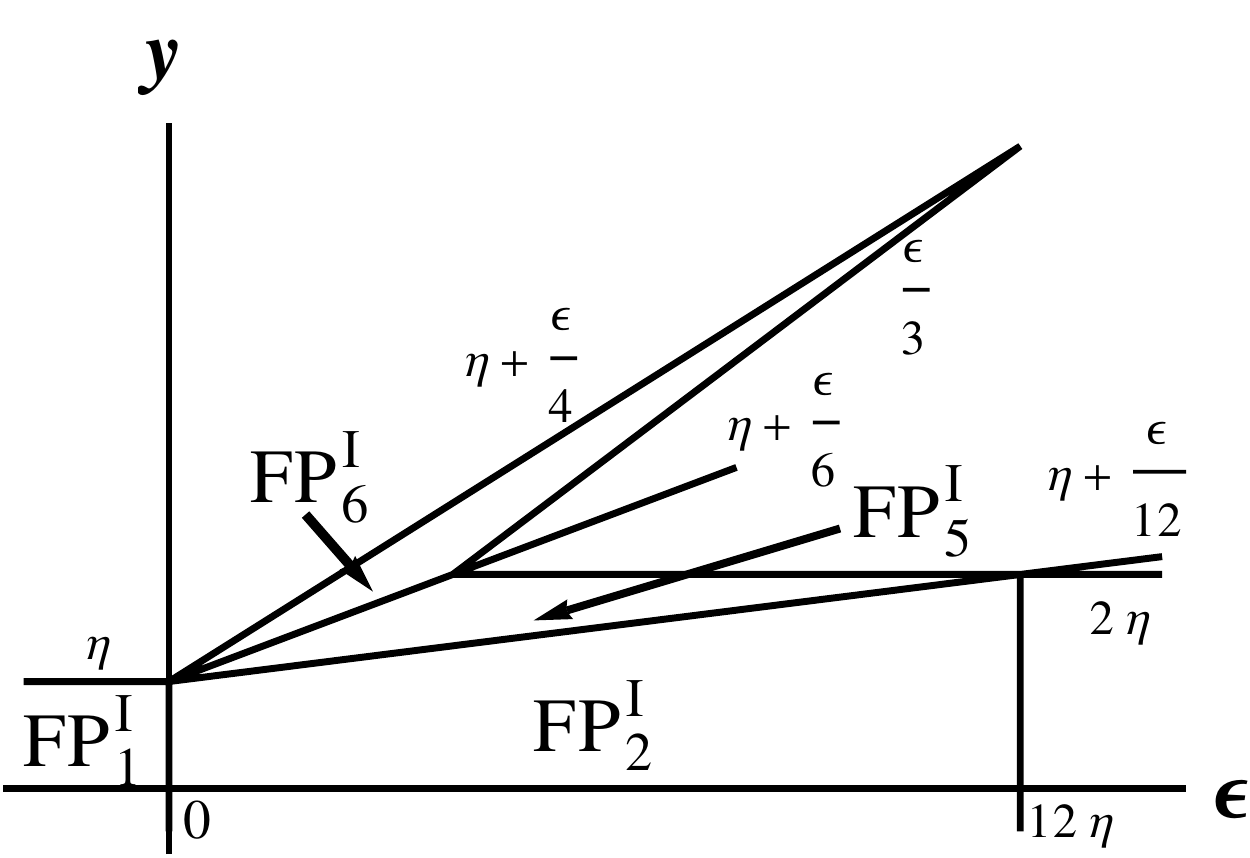} 
  \caption{Qualitative structure of the phase structure for the rapid change model
  in the $(\eps,y)$-plane.}
  \label{fig:stab_rchm}
\end{figure}

{\subsection{Rapid change model} \label{subsec:rchm}}
The rapid change model \cite{Ant00} is characteristic by  $u_1^*=0$ which, 
having in mind the replacement (\ref{eq:new_charges}), corresponds to velocity
propagator (\ref{eq:potential_v}) with short range correlations in time.
For this case seven FPs were found. Out of them only four (\fp{I}{1}, \fp{I}{2}, 
\fp{I}{5} and \fp{I}{6}) are IR stable. Coordinates of all fixed points
are listed in Table \ref{tab:rchm}. As we can see the coordinates
depend only on the difference $y-\eta$, which confirms previous expectations \cite{Ant99,Ant00}. 
\begin{table}[!ht]
  \centering
  \setlength\extrarowheight{4pt}
  \begin{tabular}{|c||c|c|c|c|}
    \hline
    \fp{I}{} & $g_1$ & $g_2$ & $u_2$ & $a$ \\
    \hline
    \hline
    \fp{I}{1} & $0$ & $0$ & NF & NF \\
    \hline
    \fp{I}{2} & $0$ & $\frac{2}{3}\eps$ & $0$ & $0$ \\
    \hline
    \fp{I}{3} & $4(y-\eta)$ & $0$ & $0$ & NF \\
    \hline
    \fp{I}{4} & $4(\eta-y)$ & $0$ & $\frac{1}{2}$ & $0$\\
    \hline
    \fp{I}{5} & $\frac{1}{3}[12(y-\eta)-\eps]$ & $\frac{2}{3}\eps$ & $0$ & $0$ \\
    \hline
    \fp{I}{6} & $\eps +4(\eta-y)$ & $\frac{2}{3} \eps$ & $\frac{\eps +6(\eta-y)}{3[\eps+4(\eta-y)]}$ & $0$ \\
    \hline
    \fp{I}{7} & $\eta-y$ & $2(y-\eta)$ & $1$ & $-6+ \frac{2\eps}{y-\eta} $\\
    \hline
  \end{tabular}
  \caption{Coordinates of the fixed points for the rapid change model.}
  \label{tab:rchm}
\end{table}
Here NF stands for Not Fixed, i.e., the corresponding value
of a charge coordinate could not be unambiguously determined. In that case the given FP,
rather than to a point, corresponds to the whole line of FPs. From the explicit
expression for $\beta_a$ in (\ref{eq:final_beta}) we can draw a conclusion about points for which ${a}^*=0$.
 For them this relation is exact and is fulfilled to all orders in a perturbation theory.
 The schematic depiction of the phase space structure can be found in Fig. \ref{fig:stab_rchm}.
The phase boundaries for the \fp{I}{6} can be obtained only in the numerical way and therefore they are not included. 

\begin{figure}[h!]
  \centering
  \includegraphics[width=6cm]{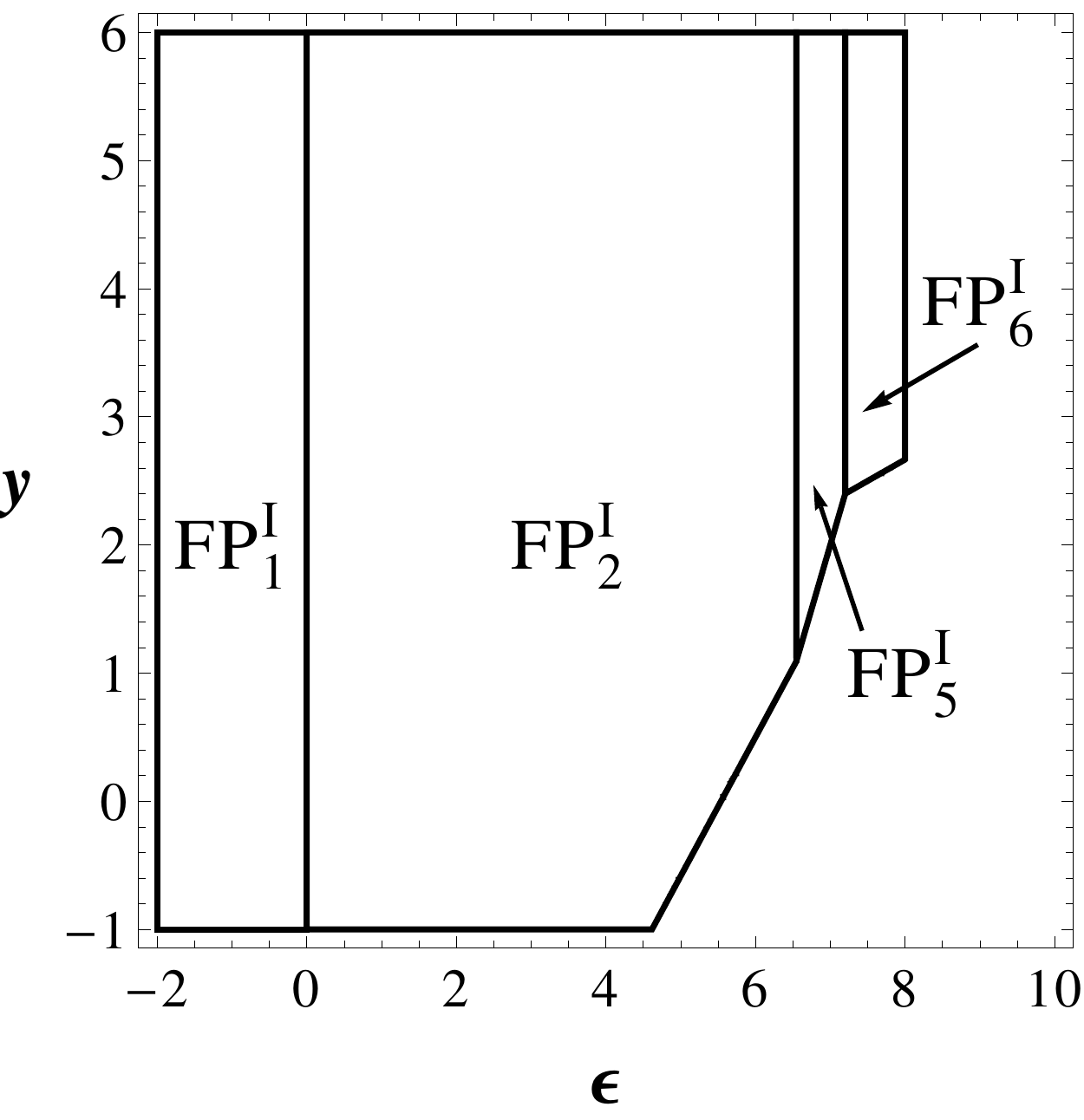}
  \caption{Phase portrait for the percolation process in the presence of
  thermal fluctuations in the $(\eps,y)-$plane. The notations for the fixed points agrees with that
  of rapid-change model in Table \ref{tab:rchm}.}
  \label{fig:thermal}
\end{figure}

\fp{I}{1} corresponds to the free (Gaussian) theory for which all interactions are irrelevant 
and ordinary perturbation theory is applicable. However, existence of such point is necessary
condition for the RG approach \cite{Vasiliev}.
For the \fp{I}{2} the correlator of the velocity field is irrelevant and 
this point describes a standard DP universality class \cite{JanTau04}.
Remaining two fixed points represent nontrivial regimes, for which velocity
fluctuations as well as percolation interaction are relevant.
 In Fig. \ref{fig:stab_rchm} we observe that the realizability of the regime \fp{I}{5} 
crucially depends on the non-zero value of $\eta$.
{\subsection{Thermal fluctuations} \label{subsec:thermal}}
Now we analyze a special case of the rapid-change model, which describes
thermal fluctuations \cite{FNS}. They
are characterized by quadratic dispersion law and in our choice of
velocity correlator (\ref{eq:kernelD}), an additional condition 
\begin{equation}
  \eta = 6 +y - \eps   
  \label{eq:cond_thermal}
\end{equation}
has to be met. 
A phase structure in the plane $(\eps,y)$ is depicted in
Fig. \ref{fig:thermal}. For physical space dimensions 
$d=3\mbox{ }(\eps=1)$ and $d=2\mbox{ }(\eps=2)$ the only stable regime is that of pure DP.
The nontrivial regimes \fp{I}{5} and \fp{I}{6} are realized only in the nonphysical
region for large values of $\eps$. This numerical result confirms our
previous expectations \cite{AntKap08,AntKap10}. 
It was pointed out \cite{HH00} that
genuine thermal fluctuations can change IR stability of the given universality class. 
However, this is not realized for the percolation process itself.
{\subsection{Frozen velocity field} \label{subsec:frm}}
Frozen velocity limit is obtained for the choice of $u_1=1$. 
In the formulation of an advection of density field
 \cite{Ant00}  it corresponds to the model of random walks in a random environment with long-range
 correlations \cite{HonKar88}.
In this case
five fixed points can found and their coordinates are listed in Table \ref{tab:frozen}. Only
the points \fp{II}{1}, \fp{II}{2} and \fp{II}{4} are IR stable.
The fixed point \fp{II}{1} describes the free (Gaussian) theory. It is stable in the region
\begin{equation}
   y<0,\quad \eps < 0, \quad \eta < 0.
   \label{eq:free_frozen} 
\end{equation}

\begin{figure}[h!]
   \centering
   \begin{tabular}{ c c}
     \includegraphics[width=6cm]{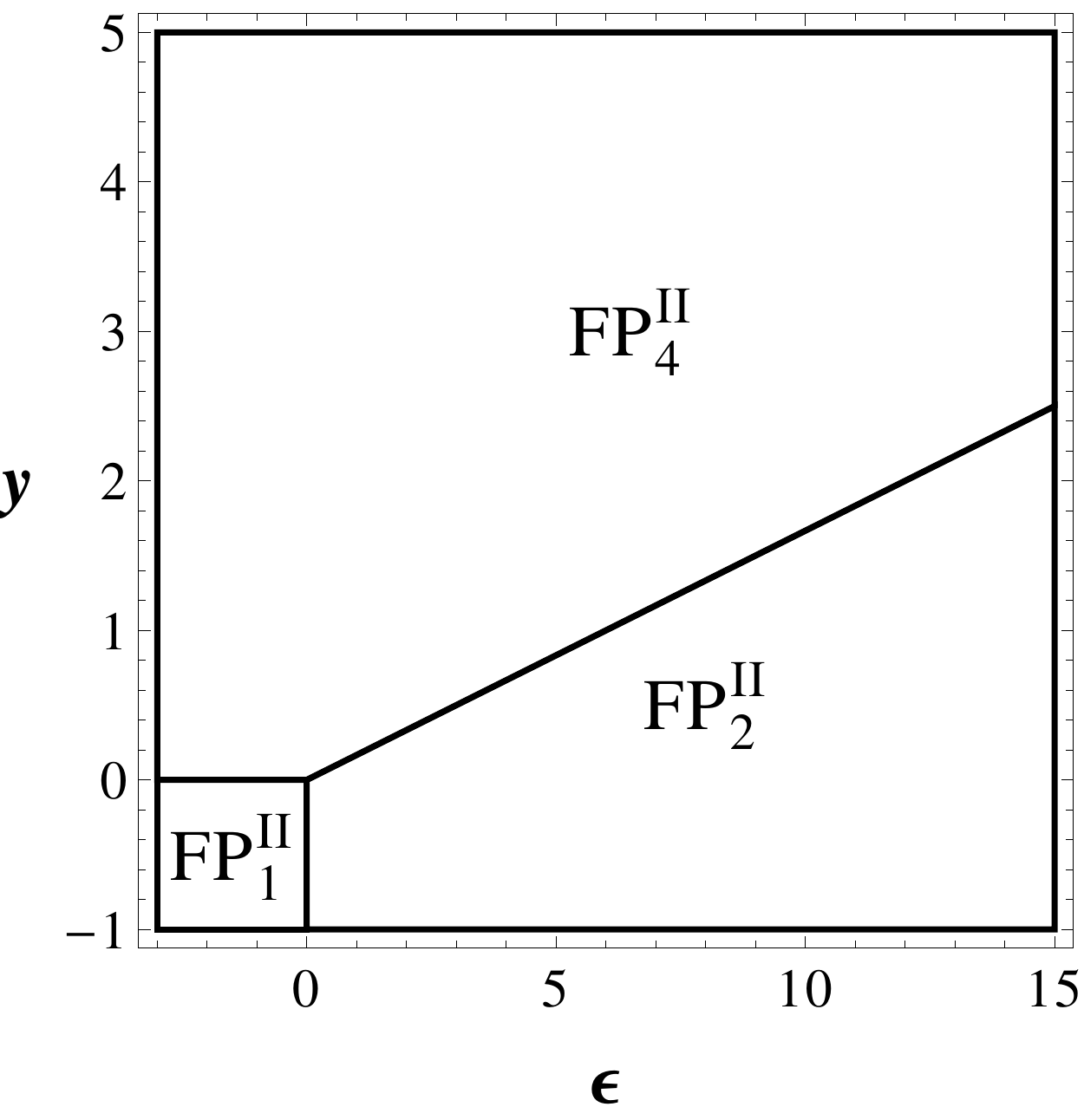}
     &
     \includegraphics[width=6cm]{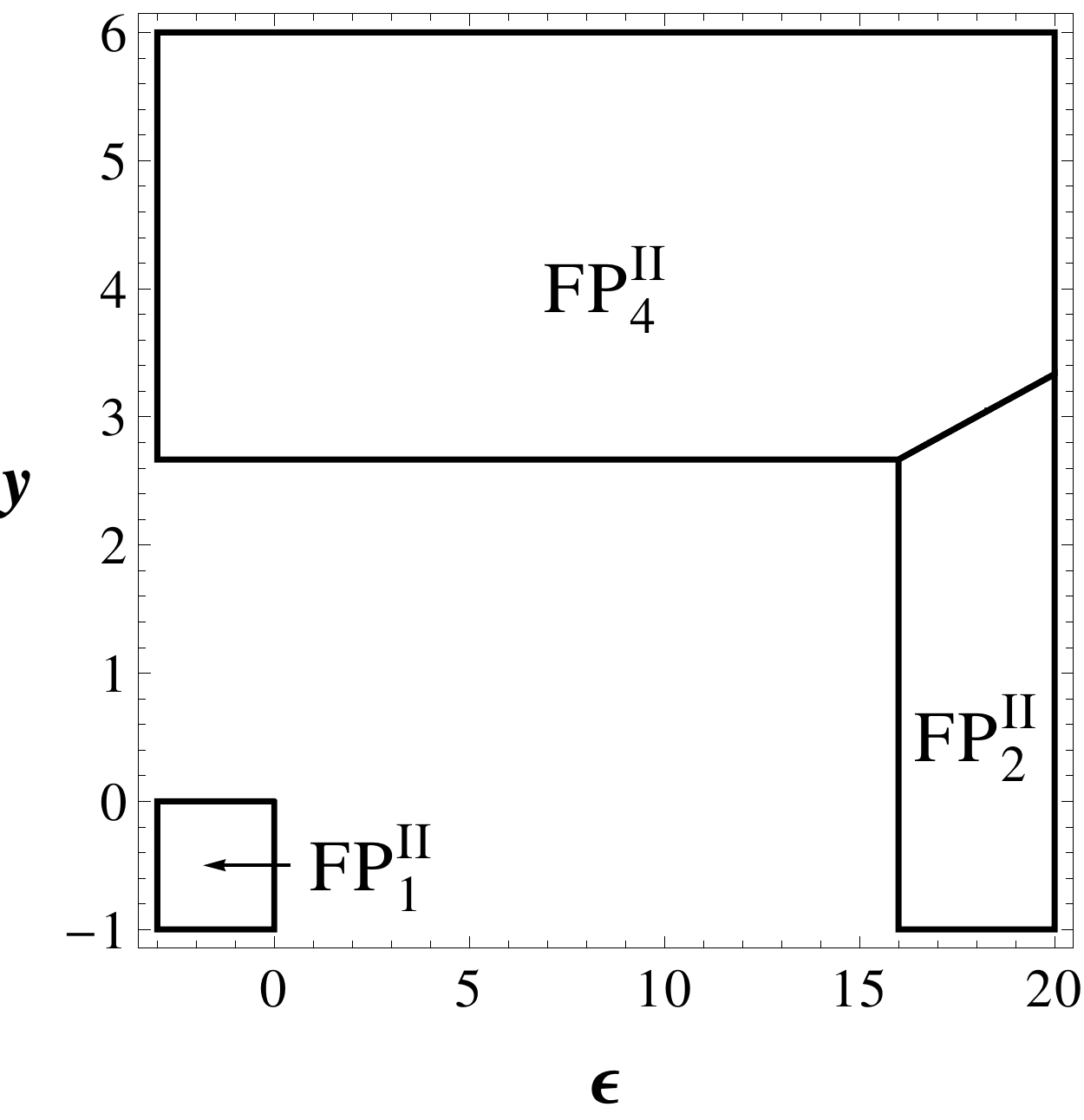}
   \end{tabular}
   \caption{On the left picture region of stability for $\eta=0$ in the plane $(\eps,y)$ is depicted 
   and on the right picture phase structure for the choice $\eta=4/3$.}
   \label{fig:frozen}
\end{figure}

\begin{table}[!ht ]
  \centering
  \setlength\extrarowheight{7pt}
  \begin{tabular}{| c||c|c|c|c |}
    \hline
    \fp{II}{} & $g_1$ & $g_2$ & $u_2$ & $a$ \\
    \hline\hline
    \fp{II}{1} & $0$ & $0$ & Not Fixed & Not Fixed \\
    \hline
    \fp{II}{2} & $0$ & $\frac{2}{3}\eps$ & $0$ & $0$ \\
    \hline
    \fp{II}{3} & $\eps-y$ & $4 y $ & $1$ & $\frac{5 y - \eps }{\eps - y}$ \\
    \hline
    \fp{II}{4} & $4( 6 y - \eps )$ & $4 y$ & $\frac{4 \eps - 25 y + \sqrt{-8 \eps y + 49 y^2}}{8(\eps - 6y)}$ & $0$ \\
    \hline
    \fp{II}{5} & $4( 6 y - \eps )$ & $4 y$ & $\frac{4 \eps - 25 y - \sqrt{-8 \eps y + 49 y^2}}{8(\eps - 6y)}$  & $0$ \\
    \hline
    \end{tabular}
    \caption{Coordinates of the fixed points for the frozen velocity ensemble.}
    \label{tab:frozen}
\end{table}
For the \fp{II}{2}
the velocity field is irrelevant and the only 
relevant interaction is the nonlinearity of the percolation process. This FP is stable in
 the region
 \begin{equation}
   \eps > 6y, \quad \eps > 0,\quad \eps > 12\eta.
   \label{eq:DP_frozen}
\end{equation}
 \fp{II}{4} embodies a nontrivial regime for which both velocity and
percolation interactions are relevant. The regions of stability for the \fp{II}{1} and \fp{II}{2} 
are depicted in Fig. \ref{fig:frozen}. Because for these two points
the velocity field could be effectively neglected, it directly follows that
 given boundaries could not depend on the value of parameter $\alpha$. The stability region
of \fp{II}{4} can be computed only numerically and it turns out that it depends on $\alpha$. 
From Fig. \ref{fig:frozen} we observe that the correlation parameter $\eta$ crucially affects
boundaries between \fp{II}{2} and \fp{II}{4}.
{\section{Conclusions} \label{sec:conclu}}
In this paper we have studied percolation spreading in the presence of
 irrotational velocity field with long-range correlations. The coarse grained model was
 formulated and multiplicative
renormalizability of the field theoretic model was discussed in detail. 

We have found that depending on the values of a spatial dimension $d=4-\eps$, scaling
exponents $y$ and $\eta$, describing statistics of velocity fluctuations, the
 model exhibits various universality classes.
Some of them are already well-known: the Gaussian (free) fixed point,
 a directed percolation without advection and a passive scalar advection. The remaining
 points correspond to new universality classes, for which an interplay
between advection and percolation is relevant.

It was shown \cite{GV99} that anomalous scaling behavior is destroyed when $\alpha$ and $y$ are large
enough. Therefore only relatively small values of
$\alpha$ are allowed ($\alpha \ll 1$) in our model. They correspond to small fluctuations of the density 
$\rho$, what is tacitly supposed in our investigation. In other words, it is assumed that the
 velocity of the fluid is much smaller than the velocity
of the sound in the system (the Mach number $\mathrm{Ma} \ll 1$). 
 Nevertheless we believe that a qualitative picture for
large values of compressibility should remain the same.
 A possible further investigation should take into account additional effects such as feedback on the dynamics of the
advecting field, anisotropies or broken mirror symmetry.

The work was supported by VEGA grant No. $1/0222/13$ 
 of the Ministry of Education, Science, Research and Sport of the Slovak Republic. N.~V.~A. and
  A.~S.~K. acknowledge Saint Petersburg State University for Research Grant No. 11.38.185.2014.
  A.S.K. was also supported by the grant 16-32-00086 provided by the Russian
Foundation for Basic Research.

\end{document}